\documentclass[twocolumn,showpacs,preprintnumbers,amsmathamssymb]{revtex4}

\usepackage{graphicx}
\usepackage{epsfig}
\begin{document}

%\DeclareGraphicsExtensions{.eps, .jpg} 

%%%%%%%%%%%%%%%%%%%%%%%%%%%%%%%%%%%%%%%%%% 

\bibliographystyle{prsty}

\title{Pressure and alloying effects on the metal to insulator transition in NiS$_{2-x}$Se$_{x}$ studied by infrared spectroscopy} 
 
\author {A.~Perucchi$^{1,2}$, C.~Marini$^{2}$, M.~Valentini$^{2}$, P.~Postorino$^{2}$, R.~Sopracase$^{2}$, P.~Dore$^{2}$, P.~Hansmann$^{3,4}$, O.~Jepsen$^{3}$, G.~Sangiovanni$^{3,4}$, A.~Toschi$^{3,4}$, K.~Held$^{4}$, D.~Topwal$^{5}$, D.D.~Sarma$^{6}$, and S.~Lupi$^{2}$} 
\affiliation{$^1$ Sincrotrone Trieste S.C.p.A., Area Science Park, I-34012 Basovizza, Trieste, Italy}\ 
\affiliation{$^2$ CNR-INFM COHERENTIA and Dipartimento di Fisica, Universit\`a di Roma "La Sapienza", Piazzale Aldo Moro 2, I-00185 Roma, Italy}\ 
\affiliation{$^3$ Max-Planck Institut f\"ur Festk\"orperforschung, Heisenbergstrasse 1, D-70569 Stuttgart, Germany}\ 
\affiliation{$^4$ Institute for Solid State Physics, Vienna University of Technology, 1040 Wien, Austria}\ 
\affiliation{$^5$ International Centre for Theoretical Physics (ICTP), Strada Costiera 11, 34100 Trieste, Italy}\ 
\affiliation{$^6$ Centre for Advanced Materials, Indian Association for the Cultivation of Science, Jadarpur, Kolkata 70032, India and \\ Solid State and Structural Chemistry Unit, Indian Institute of Science, Bangalore 560012, India}\  
\date{\today}

\begin{abstract} 
The metal to insulator transition in the charge transfer NiS$_{2-x}$Se$_x$ compound has been investigated through infrared reflectivity. 
Measurements performed by applying pressure to pure NiS$_2$ (lattice contraction) and by Se-alloying (lattice expansion) reveal that in both cases an anomalous metallic state is obtained. 
We find that optical results are not compatible with the linear Se-alloying vs Pressure scaling relation previously established through transport, thus pointing out the substantially different microscopic origin of the two transitions.
 
 \end{abstract} 
\maketitle 
  
Understanding the physics of strongly correlated systems is one of 
the most challenging tasks of condensed matter research 
\cite{imada98}. Besides displaying extremely interesting physical 
behavior, their sensitivity to small changes in external parameters 
makes them highly appealing for future technological applications. 
That sensitivity is attributed to the small value of the electron 
bandwidth in comparison with other relevant energy scales as the 
electron correlation $U$ or the charge transfer (CT) energy gap. 
The independent electron approximation breaks down and 
materials at half filling can be insulators, contrary to the 
prediction of band theory. 
 
\begin{figure}[h] 
   {\hbox{\psfig{figure=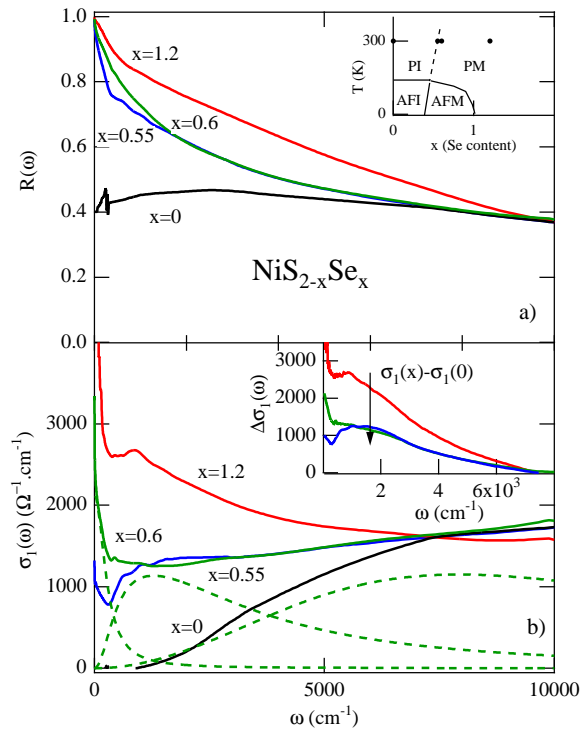,width=8cm}}} 
     \caption{(Color online) a) Optical reflectivity of NiS$_{2-x}$Se$_x$ for $x=0$, 0.55, 0.6, 1.2 at ambient conditions. Inset: Phase diagram of NiS$_{2-x}$Se$_x$ \cite{imada98}. Black dots correspond to the samples measured in this work. b) Optical conductivities from KK transformations. Thick dashed lines represent the DL fit of the $x=0.6$ sample. Inset: Difference $\Delta\sigma_1=\sigma_1(x)-\sigma_1(x=0)$ spectra. 
     } 
\label{fig1} 
\end{figure}

The cubic pyrite NiS$_2$, which is a CT insulator following the 
Zaanen-Sawatsky-Allen classification scheme \cite{zaanen85}, is 
considered, together with vanadium sesquioxide V$_2$O$_3$, a 
text-book example of strongly correlated materials. NiS$_2$ attracts 
particular interest as it easily forms a solid solution with 
NiSe$_2$ (NiS$_{2-x}$Se$_x$), which, while being iso-electronic and 
iso-structural to NiS$_2$, is nevertheless a good metal. A metal to insulator transition (MIT), 
induced by Se alloying, is observed at room temperature (T) for 
$x\approx0.6$, and a magnetic phase boundary from an 
antiferromagnetic to a paramagnetic metal is found at low 
T at about $x=1$ (see the inset of Fig.\ref{fig1}a) \cite{imada98}. 
 An alternative way to induce a metallic state in NiS$_2$ is 
applying a hydrostatic pressure (P). Following Mott's original idea \cite{Mott}, this technically challenging procedure offers the unique opportunity to continuously tune the bandwidth, 
without introducing impurities or disorder. High-P techniques 
have indeed been used in the past few years to investigate the dc 
transport properties of NiS$_{2-x}$Se$_x$ 
\cite{miyasaka00,niklowitz06}, and a P induced MIT has been 
observed in pure NiS$_2$ for P $>$ 4 GPa. 
 
Infrared reflectivity, in particular under pressure,  is a very suitable probe to address the physics of strongly correlated systems. The investigation of the T-dependent optical properties of V$_2$O$_3$ and their theoretical explanation in terms of coherent and incoherent excitations around the Fermi energy (E$_F$) represents one of the most compelling successes of the dynamical mean field theory (DMFT)  \cite{rozenberg95, kotliar06, baldassarre08}. However, with few remarkable exceptions \cite{okimoto95, congeduti, postorino, pashkin06}, infrared investigations of the MIT in CT insulators are still rare, and a thorough optical study of NiS$_{2-x}$Se$_x$ $vs$ Se-alloying and applied-P is completely lacking to the best of our knowledge. 
In this paper we fill this gap, presenting room-T reflectivity measurements over a broad spectral range on 4 compounds ($x=0$, 0.55, 0.6, 1.2) of the NiS$_{2-x}$Se$_x$ series together with optical measurements as a function of P on pure NiS$_2$. 
Experimental data are compared with local density approximation (LDA) calculations, and the resulting scenario for the two MITs is finally depicted.
 
The ambient-P nearly normal incidence reflectivity $R(\omega)$ has been measured between 50 and 
35000~cm$^{-1}$ on 
well characterized high density pellets of NiS$_{2-x}$Se$_x$ 
\cite{sarma03,sample}. An {\it in situ} evaporation technique was used to measure the reference. The high-P study has 
been performed using a diamond anvil 
cell (DAC). A small piece of NiS$_2$ was loaded inside the gasket hole together with KBr as hydrostatic 
medium. Great care was taken to obtain a clean sample-diamond interface where reflectivity spectra, 
$R_{sd}(\omega)$, have been measured  \cite{sacchetti07}.  
The measurement was performed at the high brightness infrared 
synchrotron radiation source SISSI@Elettra (Trieste) \cite{lupi07}. 
Further  details on the measurement procedures are reported elsewhere \cite{baldassarre07, 
arcangeletti}. 
 
The $R(\omega)$ of NiS$_2$ at ambient-P, shown in 
Fig.\ref{fig1}a, is nearly flat from 50 to 10000~cm$^{-1}$ except 
for weak phonon contributions at 260 and 290~cm$^{-1}$. 
On increasing the Se-content, $R(\omega)$ is progressively 
enhanced at low frequencies, characteristic of a metallic behavior. 
The real part of the optical conductivity $\sigma_1(\omega)$ (Fig.\ref{fig1}b) has been determined through Kramers-Kronig (KK) 
transformations. To this end, standard extrapolation procedures were 
adopted at both high and low frequency \cite{wooten,dressel}. 
 
The optical conductivity of NiS$_2$ is strongly 
depleted at low frequency showing the CT gap (evaluated at Full-Width-Half-Maximum of the absorption). This is consistent with previous optical 
measurements \cite{kautz72, miyasaka00} at 
about 4000~cm$^{-1}$. On increasing the Se-content $x$, a large 
amount of spectral weight (SW)  is 
transferred from high to low frequency through an
isosbestic point around 8000~cm$^{-1}$. This suggests the main 
role played in the MIT by electronic correlations \cite{imada98}. 
As it is better highlighted by the 
$\Delta\sigma_1=\sigma_1(x)-\sigma_1(x=0)$ difference spectra in the 
inset of Fig.\ref{fig1}b, the low energy contribution to the $\sigma_1(\omega)$ is made up of 
two well distinct terms: one broad mid-IR band peaked around 2000~cm$^{-1}$ and extending up to nearly 8000~cm$^{-1}$, and a 
sharp contribution below 500~cm$^{-1}$.
In analogy with spectra of metallic V$_2$O$_3$ 
\cite{rozenberg95, baldassarre08},  the narrow peak is attributed to quasi-particle (QP) coherent excitations around $E_F$, while the mid-IR term is associated to optical 
transitions from the QP peak to the upper and lower Hubbard bands. 
This scenario has been confirmed by fitting 
the $\sigma_1(\omega)$ curves through a Drude-Lorentz (DL) model. Data can be described 
by a Drude term plus two Lorentzian oscillators. The Drude and the low energy oscillator 
(centered around 2000~cm$^{-1}$) describe the coherent and the mid-IR excitations around $E_F$, 
while the remaining oscillator at 10000~cm$^{-1}$ mimics the CT and Hubbard transitions. The fitting components are reported as thick dashed 
lines in Fig.\ref{fig1}b for $x=0.6$. 

%<<<<<<<<<<<<<<<<<<<<<<<< FIGURE>>>>>>>>>>>>>>>>>>>>>>>>> 

\begin{figure}[h] 
   {\hbox{\psfig{figure=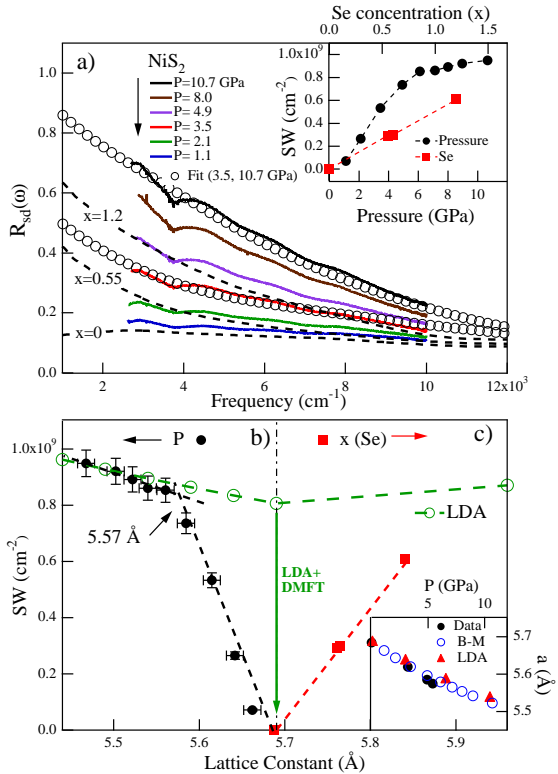,width=8cm}}} 
     \caption{(Color online) a) $R_{sd}(\omega)$ measured for NiS$_2$ at high-P (thick solid lines) 
     (open symbols show DL fits) and calculated $R_{sd}^{cal}(\omega)$ at sample-diamond interface for NiS$_{2-x}$Se$_x$ at selected $x$ (dashed lines). The kink in $R_{sd}(\omega)$ at about 3700~cm$^{-1}$  is instrumental. Inset: QP spectral weight $vs$ P (bottom) and $x$ (top); top and bottom scales are chosen consistently with a scaling factor $f\approx0.14$/GPa (see text). Lower panels: QP spectral weight (see text)  $vs$ lattice constant $a$ for NiS$_2$ (b) and NiS$_{2-x}$Se$_x$ (c). Green circles are square plasma frequency values calculated with LDA (see text). A rescaling factor of 1.25 has been used for the comparison with the experimental data. The dashed-dotted vertical line marks the $a$ value for NiS$_2$ at ambient conditions. Inset: lattice parameter $vs$ P. Experimental data from Ref.\onlinecite{fujii} (solid circles), calculated values using the B-M equation (open circles) and LDA (solid triangles).} 
\label{fig2} 
\end{figure}

We turn now to the high-P 
measurements on NiS$_2$. The reflectivity at the sample-diamond 
interface, $R_{sd}(\omega)$ (thick solid lines), is shown in Fig.\ref{fig2}a. The strong   
two-phonon diamond absorption provides reliable data  
only above 2000~cm$^{-1}$. 
On increasing the pressure, $R_{sd}(\omega)$ is progressively enhanced at low-frequency showing an overdamped behavior (similarly to $R_{sd}^{cal}(\omega)$ obtained on varying $x$), as a signature for a correlated bad metallic state. At high-frequencies all 
$R_{sd}(\omega)$ converge above 
10000~cm$^{-1}$. In order to evaluate the accuracy of our 
high-P measurements, we compare the $R_{sd}(\omega)$ data to the expected reflectivity at a sample-diamond interface, $R_{sd}^{cal}(\omega)$, calculated by using a procedure previously introduced \cite{baldassarre07,arcangeletti}. 
The calculated 
$R_{sd}^{cal}(\omega)$ for NiS$_2$ (dashed lines in Fig.\ref{fig2}a) is in good 
agreement with $R_{sd}(\omega)$ measured in the DAC at the lowest 
pressure (1.1 GPa) being both nearly flat and with a value $\approx 20 \%$ over the whole frequency range.  
The same procedure has been applied 
to NiS$_{2-x}$Se$_x$ ($x=0.55$, 0.6, 1.2) compounds and the  
resulting  $R_{sd}^{cal}(\omega)$ are shown in Fig.\ref{fig2}a for the sake of comparison. 
We then tried to fit the $R_{sd}(\omega)$ measured $vs$ P for NiS$_2$ within the same 
DL framework previously used for NiS$_{2-x}$Se$_x$ at ambient-P.  Although a 
certain degree of arbitrariness remains in fitting the data over a restricted spectral range, a reliable description of the $R_{sd}(\omega)$ at any P is obtained by the sum of a Drude and a mid-IR term plus a high-frequency oscillator kept constant at all pressures. This fit provides a robust estimate of the quasi-particle SW, defined by the sum of the Drude and of the mid-IR intensities. This sum remains nearly unchanged by varying the fitting parameters over realistic ranges.

The microscopic mechanisms 
inducing the P- and Se-MITs are further  investigated by studying the  
quasi-particle SW as a function of the cubic lattice parameter $a$. 
The lattice is expanded by Se-alloying \cite{fujimori01,kwizera80}  
whereas it is compressed by pressure \cite{fujii}. 
The $x$- and the P-dependence (up to $\approx$ 5 GPa) of $a$ have 
been obtained from Ref.\cite{kwizera80} and  Ref.\cite{fujii} 
respectively. Data at higher P have been estimated 
using the procedure developed in Ref.\cite{sacchetti07}. Through the 
specific heat results of Ref.\cite{yao97}, which provide a Debye 
frequency $\omega_D\approx 350$~cm$^{-1}$ well comparable with the 
NiS$_2$ phonon frequencies, we obtain a sound velocity $v_s\approx4300$ m/s. As the density 
of NiS$_2$ is $\rho=4455$ kg/m$^3$, the Bulk modulus results 
$B_0=\rho\cdot v_s^2\approx 83$ GPa. Assuming 
$B(P)=B_0+B'P$, $a(P)$ is finally given by the Birch-Murnagham 
(B-M) equation \cite{murnaghan44} 
\begin{equation} 
a(P)=a(0)*\left[1+\frac{B'}{B_0}*P\right]^{-1/3B'} \label{murnaghan} 
\end{equation} 
where $B'$ normally ranges between 4 and 8 \cite{jiuxun05}.
The experimental $a(P)$ data, the values estimated from Eq.\ref{murnaghan} and those from LDA calculations \cite{LDA1} are in a very good agreement as shown in the inset of Fig. \ref{fig2}c. 
%<<<<<<<<<<<<<<<<<<<<<<<< FIGURE>>>>>>>>>>>>>>>>>>>>>>>>> 
 
%<<<<<<<<<<<<<<<<<<<<<<<< FIGURE>>>>>>>>>>>>>>>>>>>>>>>>> 
The quasi-particle SW, shown for pure NiS$_2$ at working P in Fig.\ref{fig2}b
and for NiS$_{2-x}$Se$_x$ at different Se-contents in Fig.\ref{fig2}c, reveals a 
striking non-monotonic behavior as a function of $a$. Its slow continuous increase for $a < 5.57$ \AA\  (i.e. at 
the highest values of P) reflects the progressive enhancement of the  
kinetic energy due to the applied P and corresponds to a nearly-complete metallization of NiS$_2$.
For $a > 5.57\,$ \AA\  up to $a_{eq}\approx5.68$ 
\AA\ (namely the lattice parameter corresponding to NiS$_2$ at ambient conditions), 
correlation effects get larger and the SW drops 
rapidly to zero as a consequence of the Mott transition. 
On further increasing $a$ above $a_{eq}$ due to the Se-alloying, the SW 
(Fig.\ref{fig2}c) restarts to increase, 
owing to the onset of the Se-induced MIT. 
 
\begin{figure}[h] 
  {\hbox{\psfig{figure=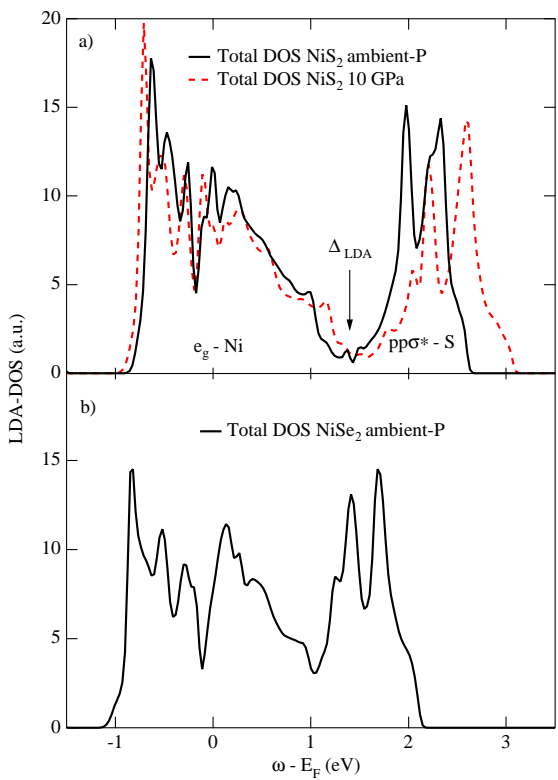,width=8cm}}} 
     \caption{(Color online): a) Total LDA-DOS for NiS$_2$. Solid line represent the DOS at ambient P, dashed lines at 10 GPa. b) Total LDA-DOS for the NiSe$_2$ compound. In the case of NiS$_2$, the 10 GPa DOS can be rescaled by a factor 0.88 on that at ambient-$P$, resulting in an increase of $W_{e_g}/U$, keeping $W_{e_g}/\Delta_{\rm LDA}$ fixed. 
     In contrast, Se-substitution results in a decrease of $W_{e_g}/U$ and an increase of $W_{e_g}/\Delta_{\rm LDA}$ due to the shrinking of the charge-transfer gap $\Delta_{\rm LDA}$. 
     } 
\label{fig3} 
\end{figure}

Despite the opposite behavior of the lattice parameter $vs$ Se-alloying and pressure, a linear scaling factor $f\approx0.14$/GPa between $x$ and P has been formerly established from low-T dc-resistivity data \cite{miyasaka00,niklowitz06}, thus suggesting an equivalency between the two MITs. However, the same $f$ does not apply comparing the optical SW dependence on P and x. 
It is indeed clear from the inset of Fig. \ref{fig2}a that the rate of increase of SW is much larger with P than with $x$.
The breakdown at finite frequencies of the dc linear scaling between $x$
and P suggests that, while a metallic state can be obtained from NiS$_2$  both by applying P and by alloying Se, this state takes place through substantially different microscopic mechanisms, involving different redistributions in the electronic density of states.
A qualitative understanding of the two different MITs can be obtained through self-consistent TB-LMTO LDA calculations \cite{ole}, see Fig.\ \ref{fig3}. To this end,
we employed the $N$th order muffin-tin orbital (NMTO) downfolding \cite{NMTO}, and the augmented plane waves plus local orbitals (APW+lo) techniques within the framework of the Wien2K code \cite{Wien2K}.
 At ambient P the Ni $e_g$-states with a bandwidth $W_{e_g}=2.1$eV and the antibonding $pp\sigma^*$-S states are separated by a CT gap ($\Delta_{\rm LDA}$) centered around 1.5 eV. Beside this gap, a second LDA-CT gap is present between occupied $pp\pi$-S states below $E_F$ and the $e_g$-Ni states (see e.g. Ref. \cite{ove}). 
Upon applying pressure, the lattice contracts and the entire band-structure around $E_F$ is renormalized; e.g. by a factor of 1.13 at $P=10\,$GPa (dashed curve in Fig.\ \ref{fig3}a).
All  features of the DOS stay the same, in particular, 
the energy  scales, ($W_{e_g}$
and $\Delta_{\rm LDA}$, are rescaled by a factor 1.13.
Hence, the bandwidth-gap ratio $W_{e_g}/\Delta_{\rm LDA}$ remains nearly constant. On the other hand, the interaction $U$ can be assumed to be constant, 
so that $W_{e_g}/U$ increases by the factor $1.13$, triggering a bandwidth-controlled MIT (BC-MIT). 
In the case of Se-substitution, the lattice expands (instead of shrinking)
due to the larger atomic radius of Se ions.
This leads to a very complementary scenario: While the changes of the $e_g$-bandwidth $W_{e_g}$ are negligible, the CT gaps shrink (see Fig.\ \ref{fig3}b). Assuming $U$ again to be constant, the driving force for the MIT is now the 
 the reduction of the charge transfer gap ($W_{e_g}/\Delta_{\rm LDA}$ increases, as $W_{e_g}$ remains  basically unaffected).

Let us now turn back to the optical experiment in Fig.\ \ref{fig2}.
Deep inside the metallic phase, we can expect correlations to be weak
 and LDA to
give the proper answer. 
Within LDA we calculated the "square plasma frequency", which is defined as the average over the FS of the squared velocities. In Fig. \ref{fig2}b-c we compare the square plasma frequency to the experimental SW.
Because of the different changes in the bandstructure under pressure and
upon Se alloying, LDA gives non-monotonic behavior and very different slopes
in agreement with experiment, see Fig.\ \ref{fig2}b-c).
For the insulating NiS$_2$ compound and close to the phase transition,
electronic correlations are not negligible and we performed LDA+DMFT \cite{LDADMFT1}
calculations. To this end, 
the NMTO bandstructure was downfolded to effective Ni  $e_g$-states
and the correlations in these two orbitals were treated by means of DMFT.
For  $U > 3J$ ($U$ being the intra-orbital Coulomb interaction between the
Ni  $e_g$-states and $J$ the Hund coupling), the two $e_g$-orbitals split and a gap opens
for NiS$_2$. 
This insulating LDA+DMFT solution results in a very strong suppression 
of the square plasma frequency (as well as of the SW) as indicated by the 
vertical arrow in  Fig.\ \ref{fig2}b-c.

Besides important similarities between the P- and Se-dependent phase diagrams \cite{miyasaka00,fujimori01}, the present optical study
reveals that a simple linear scaling between P and $x$, as that indicated by transport, does not hold at finite frequencies.This suggests that the two MITs rely on distint microscopic mechanisms.
These mechanisms can be understood theoretically in terms of the two fundamental parameters for the MIT in a CT insulator:
Under pressure,  $W_{e_g}/\Delta_{LDA}=\mbox{const.}$ and  $W_{e_g}/U$ increases, triggering the MIT; in contrast upon alloying Se, the increase of $W_{e_g}/\Delta_{LDA}$ is responsible for the MIT, whereas $W_{e_g}/U$ even decreases.
This makes  NiS$_{2-x}$Se$_x$ under pressure an ideal system for the study of the MIT in a CT strongly correlated system.

\acknowledgments The authors acknowledge L. Baldassarre and E. 
Arcangeletti for preliminar optical measurements and M. Polentarutti 
for x-ray characterization of the NiS$_2$ compound. This work was supported by the Austrian Fonds zur  F\"orderung der  wissenschaftlichen Forschung for founding.

\newpage


\begin{thebibliography}{30}
\bibitem{sample} 
Due to the cubic NiS$_{2-x}$Se$_x$ structure, IR data on high density pellets are not affected by anisotropic effects.
\bibitem{LDADMFT1} 
The Hirsch-Fye quantum Monte Carlo simulation has been used for solving the DMFT equations.
\end{thebibliography}
\end{document}